# Lifeline: Emergency Ad hoc Network

Se-Hang Cheong[1], Kai-Ip Lee[2], Yain-Whar Si[3], Hou U-Leong[4]
Faculty of Science and Technology, University of Macau
dit.dhc@lostcity-studio.com[1], dit.heinzl@lostcity-studio.com[2], fstasp@umac.mo[3], ryanlhu@umac.mo[4]

*Abstract—* **Lifeline is a group of systems designed for mobile phones and battery powered wireless routers for forming emergency Ad hoc networks. Devices installed with Lifeline program can automatically form Ad hoc networks when cellular signal is unavailable or disrupted during natural disasters. For instance, large scale earthquakes can cause extensive damages to land-based telecommunication infrastructures. In such circumstances, mobile phones installed with Lifeline program can be used to send emergency messages by the victims who are trapped under collapsed buildings. In addition, Lifeline also provides a function for the rescuers to estimate the positions of the victims based on network propagation techniques. Lifeline also has the ability to recover from partial crash of network and nodes lost.**

## I. INTRODUCTION

Cellular networks constitute the backbones of today telecommunication industry. However, these networks could be disrupted during natural disasters, such as earthquakes, hurricane, tsunami, etc. In such situations, mobile phone users may not be able make emergency calls since cellular signal is likely to be unavailable due to the destruction of land-based network infrastructures. In this paper, we propose an approach for constructing and deploying an emergency Ad hoc network called *Lifeline*. Ad hoc networks are formed by a group of wireless mobile devices without relying on any existing network infrastructure or centralized administration [1].

The main function of Lifeline is to instantly form an Ad hoc emergency network based on WiFi signals from mobile phones of disaster victims and backup battery powered wireless routers deployed by search and rescue teams. Such system is very useful in emergency cases. For example, when a person is trapped and has no signal to make a call, he/she can use the Lifeline Apps in his mobile phone to connect to the Ad hoc networks formed by similar Lifeline installed devices. After that, the victim can send the messages to an emergency station via the wireless Ad hoc network. Lifeline also has the ability to recover from partial crash of network and nodes lost. In this paper, we show how commonly available mobile devices and routers can be used to form an Ad hoc emergency network. We also propose algorithms for locating the approximate position of the mobile devices in Ad hoc network.

We briefly review the related work in section 2. The motivation and general overview of Lifeline is discussed in section 3. The design of the Lifeline program is given in section 4. Section 5 illustrates the functions provided by the Lifeline. We summarize our ideas in section 6.

## II. RELATED WORK

In the last one decade, emergency Ad hoc networks have received abundant attention from different research communities [3][5][7][8][9][11][12]. Emergency Ad hoc networks aim at providing connectivity and copes with communication congestion even in disaster situations; while current cellular systems are capable of providing efficient and stable communications. Most importantly, Ad hoc networks can operate effectively for disaster relief operations even if infrastructure facilities are corrupted.

There are various routing protocols for Ad hoc networks as discussed in [2][10][14][15][17][20], such as BATMEN [10], AODV [14], and OLSR [2][20]. According to the study in [1][6], the performance of OLSR is more stable than BATMEN and AODV on different network situations. Therefore, we choose OLSR (Optimized Link State Routing) protocol as the Ad hoc network protocol in this work.

OLSR is an optimization of the pure link state algorithm, which is most popular link state algorithm in the open source world [20]. The extensibility of OLSR is very good as it works well with many extensions, such as link quality and fisheye-algorithm [13]. Moreover, the OLSR protocol is a pro-active routing protocol that constructs a route for data transmission by storing a routing table inside every node of the network. The routing table is updated upon the topology information, where the updates are exchanged by topology control (TC) packets. Every node in the network constructs a TC packet for their neighbors list in turn. A node is considered a neighbor if and only if it can be reached via a bi-directional link. Moreover, the control packets of OLSR are only flooded within some elected nodes (*relays*) in the network. This relays can significantly reduce the total amount of control traffic. Typically, the *relays* are those nodes that can reach their neighbors 2-hops far away.

OLSR recently studied by [4][16] that integrate localization services into the routing protocol. In addition, improving the lifetime of OLSR networks is studied in [18]. The basic idea is to elect high quality relays based on a new concept *maximum forcedness ratio*. The Ad hoc networks are shown to have longer lifetime using their election criteria.

## III. MOTIVATION

Lifeline has been designed to achieve following objectives.

1. Ability to construct a self-organizing Ad hoc emergency network based on WiFi when cellular signal is unavailable.
2. Ability to recover from partial crash of network and nodes lost.
3. Forward and receive emergency messages via the Ad hoc network.
4. Function to estimate the location of a device which sent an emergency message.

Lifeline program was first designed for mobile phones only. However, the maximum coverage of WiFi signal in Mobile phones is usually around 100 meters and mobile phones have only limited battery life. They cannot function for more than 24 hours when WiFi function is turned on. In addition, most of the mobile phone users do not carry extra batteries.

Routers have larger WiFi signal range compared to mobile phones. Therefore, we extended the Lifeline application so that WiFi enabled battery powered routers can also be used in the proposed mobile Ad-hoc network. Battery powered routers can used in following scenarios:

- Routers with backup battery can be installed with Lifeline program and can be mounted in cities or residential areas where disasters such as earthquake are likely to occur. For instance, these routers can be installed a long side with emergency exit lights in hospital, schools, or public places. When earthquake strikes and electricity is cut off, the router which had no electricity supply will be reset after a certain amount of time. After that, these routers which are running in battery mode automatically boot up to form Ad hoc networks with other nearby Lifeline installed devices.
- After the earthquakes, rescuers can airdrop or install battery powered routers in affected areas. These routers which are installed with Lifeline program automatically form Ad hoc networks or join existing ones.
- Another way of applying this network is in a skiing resort where there is a potential of having an avalanche. Routers can be installed to cover areas where avalanche is likely to strike. When there is an avalanche, routers installed with Lifeline can be turned into emergency mode for forming Ad hoc networks. Any victims buried or trapped may use their mobile phones to send emergency messages to the rescue teams.
- Apart from using this system in disaster areas, normal buildings can also be installed such battery powered routers so as to allow communication during fire or severe storms. These routers can be activated when there is no cellular signal and electricity supply.

In addition to wider coverage, battery powered routers have also other significant advantages. Lifeline can be installed in common WiFi enabled routers. They are inexpensive compared to dedicated mobile emergency cellular network stations. These routers are also extremely portable and rescue teams can quickly deploy them when an emergency case occurred.

Lifeline program allow transmission of messages across heterogeneous devises including mobile phones and routers. Messages can be transmitted from mobile phones and will be routed to an emergency station according to the protocol designed in Lifeline installed devices. Emergency station comprises a desktop/laptop computer, a WiFi enabled router, and the backend version of Lifeline program. Figure 1 illustrates the situations when messages are transmitted through heterogeneous Ad hoc networks which are supported by Lifeline.

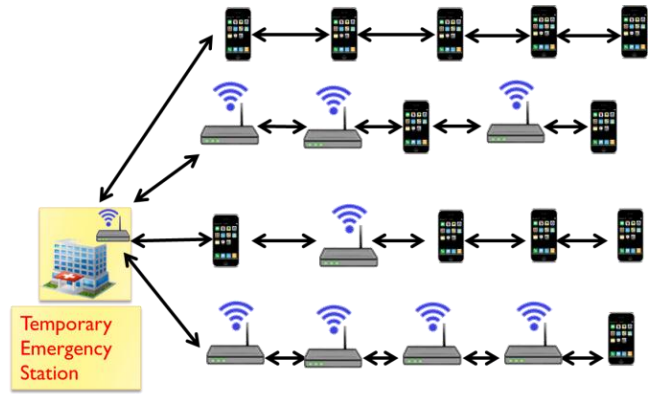

Figure 1  Ad hoc emergency networks in Lifeline.

The first scenario at the top of Figure 1 illustrates the situation where Ad hoc networks are formed by mobile phones only. The second and third scenarios depict the situation where emergency messages are transferred through a series of mixed combination of mobile phones and routers. The fourth scenario illustrates the situation where Ad hoc networks are formed by routers. Lifeline can be installed in WiFi enabled Android based mobile phone and routers, and Desktop/Laptop computers. Functions provided in Lifeline system for WiFi enabled Android based mobile phones include:

1. Forming and joining of Ad hoc networks: When cellular signal is unavailable, the mobile phone can be configured to form new Ad hoc networks or join available ones.
2. Instant message sending and receiving: Allows sending and receiving of emergency messages through Ad hoc networks. The messages will be forwarded to an emergency station.
3. Wait-for-help Timer: Shows the total time user has been waiting for help since the last emergency message is sent out.
4. One-click call for help button: In case the trapped mobile phone user is not able to type or do anything, he/she can send an emergency by pressing and holding the red button (see Figure 2).
5. Personal Information field: Personal information can be configured in advance when Lifeline program is installed. When an emergency is sent out, pre-composed personal information will be included (see Figure 2).
6. On-site environment capture: Although the exact location may not be known, the situation of the

surrounding environment of the trapped victim may give some hint to the rescue team. The user can take a photo with the mobile phone and send it with the emergency message. The screen shot of the message received at the emergency station is depicted in Figure 3.

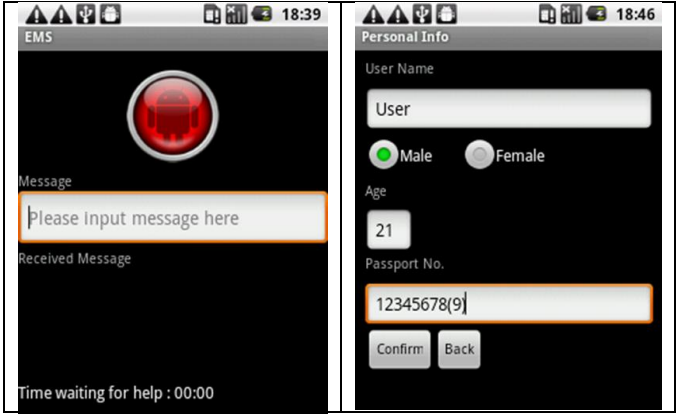

Figure 2   Emergency message editor and personal information configuration function in Lifeline for Andoid mobile phone.

Functions provided in Lifeline system for WiFi enabled routers include:
1. Auto boot up and forming of Ad hoc networks: Battery powered routers can be configured to boot up as an emergency node when electrical supply is interrupted.
2. Message forwarding: Allows forwarding of emergency messages between the victim and the emergency station.
3. Backup message module: Routers can be configured with rules to backup emergency messages received.
4. Configuring with physical location: A battery powered router can be preinstalled in residential area and they can be configured with a physical location. When an emergency case happens, the location estimation protocol can collect this information for approximating the position of the message sender.
5. Battery controlling: Through the ACPI (Advanced Configuration and Power Interface) of routers, once a router has low battery, it will pass the remaining messages to near devices and reject incoming messages gradually.

Functions provided in Lifeline system for emergency station include:
1. Instant message sending and receiving: Allows rescuers to receive and reply messages (see Figure 3).
2. Monitoring of network topology: Rescuers can monitor the topology of the nodes in the Ad hoc network (see Figure 4).
3. Approximating message sender location: Lifeline provides functions for the rescuers to approximate the location of the message sender if the nearest location of the router is known. This is the case when battery powered routers are preinstalled in buildings or routers are deployed at locations known to the rescuers.
4. Logging all victims' information: Once a victim sends messages to the emergency station, the personal information of the victim is logged for rescue mission.

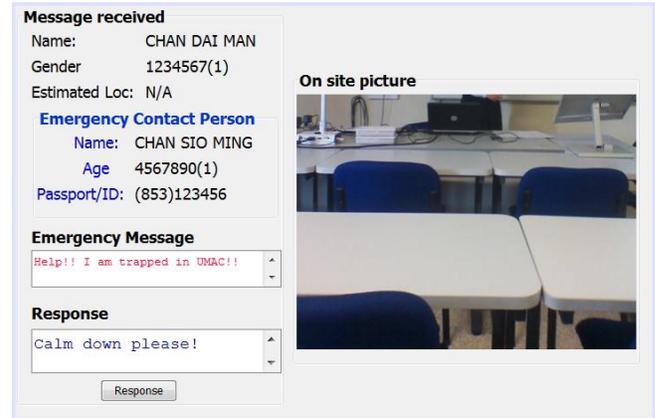

Figure 3   Reply function in emergency station.

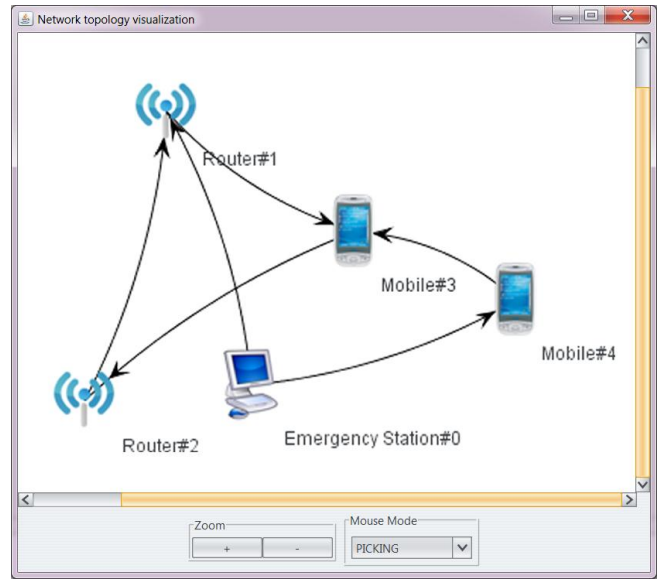

Figure 4   Network topology visualization function in emergency station.

IV. SYSTEM DESIGN

The system design of Lifeline is depicted in Figure 5. Lifeline has three major components: mobile phones, routers, and an emergency Station. Emergency station is a computer installed with a router and Lifeline backend system. Lifeline Apps is also developed for mobile phones with Android operating system. An embedded version of Lifeline is also developed for routers which support OpenWrt [19]. We use OLSR (Optimized Link State Routing Protocol) [20] as the routing protocol in our system. In the following sections, we detail the functions supported by the Lifeline system.

## A. Sending of emergency messages (Lifeline Apps in mobile phones)

Lifeline Apps in Victims' mobile phones provide functions for sending emergency messages.
- A section of IP addresses are reserved for emergency stations and these addresses are not used in other devices.
- Once the Apps is initiated, it uses HELLO and Topology Control (TC) messages to discover link state information of Ad hoc network.
- Once the network has been established, the Lifeline Apps in victim's mobile phone can send emergency messages based on OLSR routing protocol. The message will be forwarded to the next hop based on the routing path formed by nodes selected as Multipoints Relays (MPRs) [20].
- Messages are forwarded hop by hop and finally reach the emergency station.
- The reply message from emergency station to victim is transported in the same way.

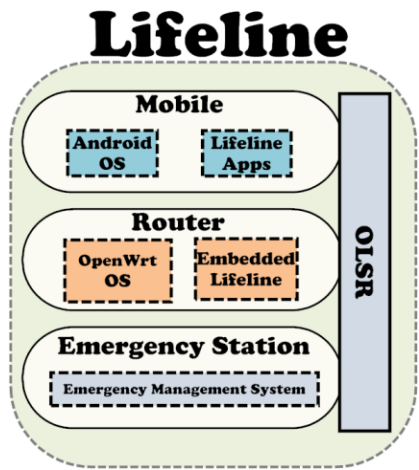

Figure 5    System design.

## B. Forward message module (Lifeline Apps for mobile phones and embedded Lifeline for routers)

Forward message module in Figure 6 accepts emergency messages from other devices and forwards them to the emergency station. Message forwarding is performed in a number of steps.

*1)* In Receive Module, we use a TCP server with port no 33333. The server is implemented in ANSI C language. In Lifeline we use XML-like message formats. The receive module accepts packets and filter out emergency messages. If the packet is an emergency message, then forward the message to the schedule module.

*2)* Schedule module determines the priority of message from the message header and the message is passed to the Forward Module. In OpenWrt System, we designed five levels of priority (0~4), each level of priority has an independent queue (see Figure 7). Messages in the higher priority queue will be forwarded to the destination first.

*3)* Forward Module stores messages in queues and sent out one by one. If the destination is unreachable then the message will be sent back to the Schedule Module and the priority of the message will be decreased by one. If an emergency message has negative priority, then the message will be swapped out to the secondary storage. When there is no emergency message left in queue Priority-0 and Priority-1 of Schedule Module, any messages which are previously swapped out to the secondary storage will be swapped in and sent to the destination again. Such swapping strategy makes sure that high priority emergency messages will be sent out as quickly as possible. Furthermore, messages in queue Priority-2, Priority-3 and Priority-4 of Schedule Module will be gradually moved into higher priority queues when messages are being sent out. That is, all messages in queue Priority-3 are moved to Priority-2, messages in queue Priority-4 are moved to queue Priority-3, and so on. When only normal priority and lower priority messages are left in the system, the system can swap in some messages from the secondary storage. Moreover, due to the limited memory available in mobile devices and routers, we can't keep all messages in the RAM. For instance, in a router, only a few megabytes of RAM are available for use for Lifeline program. In our design, if an emergency message has been swapped in and out more than twice (this threshold can be adjusted), then the message will be stored in secondary storage and will never be swapped in again. The details on performing back up of emergency messages in routers are given in the following section.

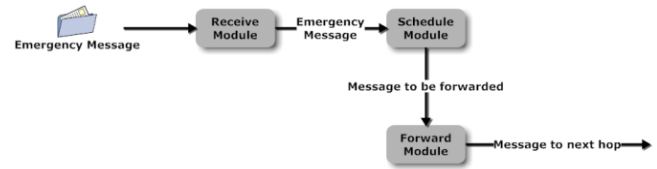

Figure 6    Message forwarding in Lifeline.

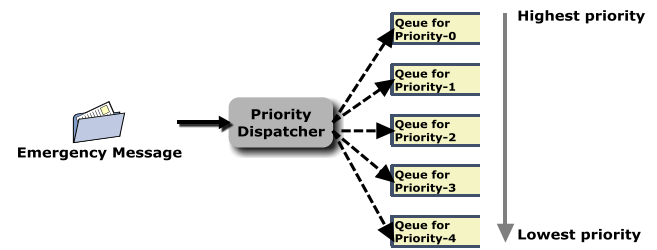

Figure 7    Priority of emergency message

## C. Backup message module (embedded Lifeline in routers)

Backup function is an extension of forward emergency messages module in routers. If the destination of message is unreachable or the battery of the router is low, then the message will be stored in permanent storage.

This function provides two advantages. Firstly, the emergency station could be unreachable due to overloading in the network or partial network crashes. Secondly, it prevents message loss which can be caused by accidental rebooting of the devices. Therefore, the system has to ensure that the messages should be stored into permanent storage before it has been sent.

In Forward Module, the system decides whether the message should be kept as a backup. The options for administrators for configuring backup function in routers for emergency messages are listed in TABLE I.

TABLE I   OPTIONS FOR PERFORMING BACKUP

| Option number | Options | Priority |
|---|---|---|
| 1 | Always backup received message. | 1 |
| 2 | Always backup message after forwarded. | 1 |
| 3 | Backup message if battery is less than *%. 0<*<=100 | 2 |
| 4 | Backup message with priority higher than *%. 0<=*<=4 | 2 |
| 5 | Backup message if the load of the current device is higher than *%. 0<*<100. | 3 |
| 6 | Backup message if the load on the source (sender) is higher than *%. 0<*<100. | 3 |

Administrators of embedded Lifeline can choose several options to configure the backup of emergency messages. In routers, higher priority option overwrites lower priority one.

For example, if we enable two options "Always backup message after forwarded" and "Backup message if the load of the current device is higher than 5%"; and the current load of system is 10%. Once an emergency message has been sent out, the message will be stored as a backup. Since "Always backup message after forwarded" has higher priority than "Backup message if the load of the current device is higher than 5%", the lower priority option is ignored.

The status of the battery can be retrieved from OpenWrt system. In addition, we can also retrieve the load of the current device (router or mobile phone). However, checking the load of the source (message sender) is problematic since the source may be far away and there may exist several devices (hops) between the source and the device which received the emergency message. Moreover, from OLSR, we can only retrieve the neighborhood information and mobile devices and routers have no a global view of the entire network. Therefore, we attach the load of the source in each message.

## D. Emergency boot module (embedded Lifeline in routers)

In the event of disasters, battery powered routers are designed to boot up in emergency mode to form Ad hoc networks. A number of alternatives are available for implementing emergency boot modules for routers.

*1) Alternative 1:* Booting up when the router detects that the power source is from the backup battery. Emergency booting process is triggered when AC power supply is interrupted and the backup battery starts power supply to device. This method relies on the function for detecting the source of power supply. However, it may cause accidental booting if the supply of AC power is unstable or if the AC power is switched off by someone.

*2) Alternative 2:* Booting up when the backup battery level drops. Since OpenWrt is also a Linux based system, we can monitor the battery status like in other distributions of Linux. Normally, the consumption of backup battery shouldn't change when the AC power is in use. If any change in the battery consumption is detected, then we can conclude that the AC power has been disrupted for a while and the emergency network should be started. This method requires the exact consumption data of the battery. The consumption information of battery from last check can be used to calculate a threshold for trigerring the boot up process.

*Consumption threshold = Battery remaining from last check - Current battery level*

Emergency booting is triggered if Consumption threshold is positive. However, this method can cause accidental booting if the battery is old or unstable. In addition, the Consumption threshold could be different in different devices and it is difficult to set a general threshold for every devices.

*3) Alternative 3:* By scanning nearby peers. In this method, the router scans its neighbors and tries to determine whether there are emergency stations or devices in emergency mode. If it founds one, then it joins the emergency network. One of the main advantages of this method is that it allows routers to provide general functions in normal situation. When a disaster occurs, then these routers can turn on OLSR protocol and form an emergency network automatically.

Figure 8 shows the scanning for nearby nodes' status by Router#1. The steps performed by Router#1 include:
1. Router#1 scans nearby WiFi devices.
2. Analyzes all information retrieved from WiFi devices.
3. Checks the signature of every device (Note: A device in emergency mode is assigned with a distinct signature).
4. Boots itself into emergency mode and connects to nearby emergency network if an emergency station or any devices in emergency mode are found.
5. Otherwise, waits for a predetermined interval, and redo from step 1 to 5.

In this method, a router may accidentally boot up into emergency mode when a nearby device in similar mode is found. The cascading effect from such situation can trigger a group of routers turned into emergency mode. In our program, we have adopted alternative 2 and 3 for implementing emergency boot up process.

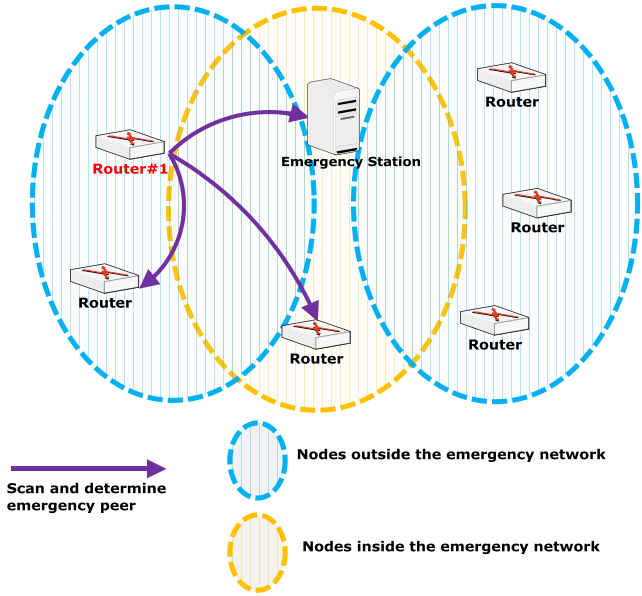

Figure 8   Scanning and determining nearby peers.

### E. Position locating (backend Lifeline program in emergency station)

When a victim is trapped under rubbles, GPS signal may not be available. Position Locating Algorithm allows rescue team and emergency station to estimate the location of the victim who sent an emergency message.

In Lifeline Ad hoc network, the estimated positions of the battery powered routers are available since they are likely to be preinstalled in buildings or deployed at known locations by the rescuers. Once an emergency network is formed, physical locations can be queried from those routers.

Based on OpenWrt installed in embedded Lifeline, we designed two fundamental protocols to exchange and retrieve physical locations between mobile phones and routers. Message transmission for position locating is depicted in Figure 9.

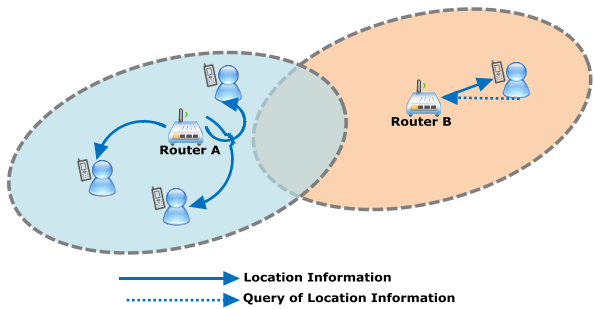

Figure 9   Two protocols for position locating algorithm.

In Figure 9, we depict the two protocols for position locating algorithm. In passive protocol for Router B's network, when a mobile phone wants to obtain the physical location, then the mobile phone can broadcast a "WHERE AM I?" to devices within N-Hops. If there is a router contains the physical location then it will reply to the portable device. If no such device near to the requester, then the requester can't determine the physical location up to N-Hops.

Another protocol is active protocol and is depicted in Router A's network in Figure 9. If a router is configured with known physical location, then the router sends its physical location to nearby devices within the network. From the OLSR routing table, a device can obtain the IDs of neighboring nodes. In the active protocol, routers can be designed to scan the network periodically based on a predefined interval to determine the joining and leaving of devices. However, this approach is inefficient and can cause overhead. An alternative approach is to detect the event change in OLSR state and once any change in the topology is detected, the router can broadcast the physical location to newly joined portable devices. This approach is more efficient than the former one.

## V. CONCLUSION

In this work, we succeed to construct our lifeline problem that enables mobile devices to automatically form emergency Ad hoc networks in different disaster situations. The Ad hoc networks are formed by a group of wireless mobile devices without relying on any existing network infrastructure or centralized administration, where the devices are communicated via OLSR protocol. In addition, we also propose an algorithm for locating the approximate position of the mobile devices in ad hoc network.

Our on-going work is to minimize the energy consumption during the messages exchange in OLSR. It is very important in our motivating examples. Moreover, it is interesting to develop a more precise algorithm for locating the victims by integrating some state-of-the-art approaches.

### ACKNOWLEDGMENT

This research is funded by University of Macau.